\documentclass[12pt,a4paper,final]{iopart}
\pdfoutput=1
\usepackage{graphicx}
\usepackage{cite}
\usepackage[breaklinks=true,colorlinks=true,linkcolor=blue,urlcolor=blue,citecolor=blue]{hyperref}
\usepackage{amsfonts, amsmath, dsfont}

\def\epp{\: .}
\def\epc{\: ,}

\def\epp{\:.}
\def\epc{\:,}

\def\rhosp{\rho_\text{sp}}

\def\limth{\lim\nolimits_\text{th}}
\def\SYY{S_{\text{YY}}}
\def\Drho{\mathcal{D}[\rho]}

\begin{document}

\title[Post-BEC-quench time evolution of the one-body density matrix ]{Analytical expression for a post-quench time evolution of the one-body density matrix of one-dimensional hard-core bosons}

\author{ J. De Nardis$^{1}$  and J.-S. Caux}
\address{$^1$Institute for Theoretical Physics, University of Amsterdam, Science Park 904,\\
Postbus 94485, 1090 GL Amsterdam, The Netherlands}
\ead{j.denardis@uva.nl}





\begin{abstract}
We apply the logic of the quench action to give an exact analytical expression for the time evolution of the one-body density matrix after an interaction quench in the Lieb-Liniger model from the ground state of the free theory (BEC state) to the infinitely repulsive regime. In this limit there exists a mapping between the bosonic wavefuntions and the free fermionic ones but this does not help the computation of the one-body density matrix which is sensitive to particle statistics. The final expression, given in terms of the difference of two Fredholm Pfaffians, can be numerically evaluated and is valid in the thermodynamic limit and for all times after the quench. 
\end{abstract}


\section{Introduction}
The time evolution of non-equilibrium isolated quantum many-body systems is one of the most intriguing problems in modern-day physics \cite{2011_Polkovnikov_RMP_83}. One of the most common out-of-equilibrium protocols is the so-called quantum quench, when the system is prepared in some initial state which is not an eigenstate of the Hamiltonian driving its unitary time evolution. This allows the system to explore a much larger part of the Hilbert space than the one associated to equilibrium properties. This leads to unusual physics as for example new steady states or pre-thermalized ones
\cite{
2006_Calabrese_PRL_96, 
2007_Rigol_PRL_98, 
2006_Rigol_PRA_74, 
rigol_dunjko_08,
2008_Barthel_PRL_100, 
2010_Cramer_NJP_12, 
2011_Cassidy_PRL_106, 
2011_Calabrese_PRL_106, 
2012_Calabrese_JSTAT_P07016, 
2012_Calabrese_JSTAT_P07022, 
2014_Bucciantini_JPA_47, 
2009_Barmettler_PRL_102, 
2010_Barmettler_NJP_12, 
2009_Rossini_PRL_102, 
2010_Rossini_PRB_82, 
2009_Faribault_JMP_50, 
2010_Fioretto_NJP_12, 
2010_Mossel_NJP_12, 
2011_Igloi_PRL_106, 
2011_Banuls_PRL_106,  
2011_Rigol_PRA_84, 
2012_Brandino_PRB_85, 
2012_Demler_PRB_86, 
2012_He_PRA_85, 
2013_He_PRA_87, 
2013_Caux_PRL_110,
2013_Mussardo_PRL_111, 
2013_Kormos_PRB_88,
2013_Mitra_PRB_87,
2014_Matta, 
2013_Heyl_PRL_110, 
2013_Pozsgay_JSTAT_P07003, 
2013_Fagotti_JSTAT_P07012, 
2013_Pozsgay_Jstat_10, 
2013_Liu,
2013_Marcuzzi_PRL_111,
Sotiriadis201452,
2014_Fagotti, 
2014_DeNardis_PRA_89, 
2014_Heyl, 
2014_Essler_PRB_89, 
2014_Bertini_Sinegordon, 
2014_Fagotti_PRB_89,
2014_Wouters, 
2014_Pozsgay_Dimer, 
2014_Pozsgay_qbosons,
Neel_Long}.

Analytical expressions for the time evolution of physical observables in quantum many-body systems are not easily computable and up to now there are only few explicit results available \cite{2006_Calabrese_PRL_96,2011_Calabrese_PRL_106,
2012_Calabrese_JSTAT_P07016,2013_Mussardo_PRL_111,1751-8121-47-40-402001,2013_Collura_PRL_110,2014_Kormos_PRA_89,2014_Bertini_Sinegordon}. Numerical simulations on the other hand are limited in the range of time due to the exponential growth of the entanglement entropy after the quench and they are in any case up to now insufficently able to treat continuous systems \cite{DMRG}. These are two of the reasons why exact analytical solutions are of fundamental importance for experimental realizations and necessary checks for numerical implementations. 
\\

The usual approach to computing the time evolution of a generic operator after a quench is to expand the initial state on a basis of eigenstates and then to sum over the whole Hilbert space. 
For some systems it is indeed possible to classify the eigenstates and have expressions for the matrix elements of some physically relevant operators. An example is given in one dimension by integrable systems \cite{GaudinBOOK,KorepinBOOK} where the Algebraic Bethe Ansatz allows to obtain matrix elements and scalar products between states even in presence of nontrivial interactions between the constituents of the system \cite{1989_Slavnov_TMP_79,1990_Slavnov_TMP_82}. However these are complicated functions of the variables parametrizing the eigenstates and it is then very hard to obtain a closed-form expression for the sum over the whole Hilbert space. 

Following the recently-introduced quench action paradigm for quenches in integrable systems \cite{2013_Caux_PRL_110,2014_DeNardis_PRA_89}, (applied also in \cite{2014_Wouters,2014_Pozsgay_Dimer,Neel_Long,2014_DeLuca}) in the thermodynamic limit we can reduce the whole sum over the eigenstates to a much more limited set of states. The logic is as follows: after the quench the system is effectively described by one of its eigenstates, the saddle point state, and the whole time evolution is governed by a restricted set of excitations on this state. The thermodynamic energies and overlaps are factorized in terms of these excitations which drastically reduces the necessary amount of information to reconstruct the time evolution of a generic operator \cite{2014_DeNardis_PRA_89}. Moreover in some cases it can give a direct indication of the velocity of propagation of the information in a quenched system \cite{Essler_LIGHT}. 
\\

In this paper we focus on the Lieb-Liniger integrable model for interacting one-dimensional bosons \cite{1963_Lieb_PR_130_1}. In particular we compute the time evolution of the one-body density matrix in the Tonks-Girardeau regime \cite{1936_Tonks_PR_50,1960_Girardeau_JMP_1,2004_Paredes_NATURE_429,2004_Kinoshita_SCIENCE_305} (hard-core bosons) after a quench from the ground state of the noninteracting regime, the BEC state. 
This observable is much less trivial than the density-density correlations even at equilibrium \cite{1964_Lenard_JMP_5} and it is directly measurable in some out-of-equilibrium experimental realizations  \cite{2007_Hofferberth_NATURE_449,2008_Hofferberth_NATPHYS_4,2012_Gring_SCIENCE_337} the same being true for its Fourier transform, which represents the (bosonic) momentum distribution function.
\\

The saddle point state after the quench from the BEC state to the Lieb-Liniger model for any final interaction strength was analysed in \cite{2014_DeNardis_PRA_89} while other aspects of the same quench are also examined in \cite{2013_Kormos_PRB_88,2014_Kormos_PRA_89,1742-5468-2014-1-P01009,2014_Matta}. In \cite{2014_DeNardis_PRA_89} the time evolution of the density-density operator is computed via the quench action approach while in \cite{2014_Kormos_PRA_89} the mapping of the density operator to the free fermion basis is used. However the same method relies only on the structure of the density operator (which is equivalent to the free fermion density in this limit) and is inapplicable to the bosonic field operator. Moreover this constitutes a step towards the computation of the time evolution after a quench to the Lieb-Liniger model with generic interaction strength.
\\

The paper is organized as follows:
in section \ref{sec2} we review the quench action logic for quenches in the Lieb-Liniger model, focusing on the post-quench time evolution of physical operators in the thermodynamic limit. We restrict then to the Tonks-Girardeau regime: in section \ref{sec3} we review the time evolution of the static density-density operator and in section \ref{sec4} we present the main result of the paper on the time evolution of the one-body function. Finally in \ref{FF_Psi} it is shown how to obtain the necessary form factors of the field-field operator in the thermodynamic limit starting from their finite size expressions given in \ref{Finite_Size}. \ref{Fredholm} reviews some properties of the Fredholm determinant and Pfaffian.


\section{The quench protocol}\label{sec2}

We consider a system of $N$ bosons on a one-dimensional ring of circumference $L$ with periodic boundary conditions and the Lieb-Liniger  \cite{1963_Lieb_PR_130_1} Hamiltonian given by (with the choice $\hbar = 2m = 1$) 
\begin{equation}\label{eq:LL_Ham}
H_{LL}=  -\sum_{j=1}^N \frac{\partial^2 }{\partial x^2_j} + 2c \sum_{j>k=1}^N \delta(x_j - x_k) \epp
\end{equation}
The coupling constant $c$ parametrizes the interaction strength. We focus here on the repulsive case $c>0$ and, in particular, on the Tonks-Girardeau regime $c = \infty$ \cite{1960_Girardeau_JMP_1}. 

The exact eigenstates of \eqref{eq:LL_Ham} are Bethe Ansatz wave functions,
\begin{equation} \label{eq:EF}
\psi\left({\boldsymbol{x}}|{\boldsymbol{\lambda}}\right) = F_{\boldsymbol{\lambda}} \sum_{P\in \mathcal{S}_N}  A_P (\boldsymbol{x}|{\boldsymbol{\lambda}}) \prod_{j=1}^N   e^{i\lambda_{P_j}x_j} \epc
\end{equation}
with $\mathcal{S}_N$ all the permutations of the set $[1,\ldots, N]$ and $ F_{\boldsymbol{\lambda}} = \frac{\prod_{j>k=1}^N (\lambda_j -\lambda_k)}{\sqrt{ N!  \prod_{j>k=1}^N\left( (\lambda_j - \lambda_k)^2 + c^2\right)}}$ , $A_P ({\boldsymbol{x}}|{\boldsymbol{\lambda}}) = \prod_{j>k=1}^N \left( 1 - \frac{ic \: \text{sgn}(x_j - x_k)}{\lambda_{P_j} - \lambda_{P_k}}\right)$. The periodic boundary conditions enforce the quantization of the rapidities $\boldsymbol{\lambda} \equiv \{\lambda_j\}_{j=1}^N$ such that they solve a set of $N$ nonlinear coupled equations, \textit{i.e.} the Bethe equations \cite{1963_Lieb_PR_130_1}
\begin{equation}\label{eq:BE_Log}
  \lambda_j =  \frac{2 \pi I_j}{L} - \frac{1}{L} \sum_{k=1}^N \theta(\lambda_j - \lambda_k) \:\:\:\:\:\:\:\:\: j = 1 , \ldots N \epc
\end{equation}
where the scattering phase shift is given by
\begin{equation}
 \theta(\lambda) = 2 \arctan(\lambda/c) \epp
\end{equation}
The quantum numbers $\boldsymbol{I} = \{I_j\}_{j=1}^N$ are mutually distinct integers (half-odd integers) for N odd (even), and they label an eigenstate $|\boldsymbol{I}\rangle$ uniquely. 
The energy $\omega_{\boldsymbol{I}} $ and momentum $P_{\boldsymbol{I}}$ of the states are given in terms of their rapidities by
\begin{align}
& P_{\boldsymbol{I}} = \sum_{j=1}^N\lambda_j \epc\\
& \omega_{\boldsymbol{I}}= \sum_{j=1}^N \lambda_j^2 \epp
\end{align}

In the limit $c\to \infty$ equations \eqref{eq:BE_Log} become the standard quantization conditions for free fermionic momenta
\begin{equation}
  \lim_{c \to \infty}\lambda_j =  \frac{2 \pi I_j}{L}  \epp
\end{equation}
In this limit there is a rigorous one-to-one  correspondence between the wave function \eqref{eq:EF} and the Slater determinant for free spinless fermions \cite{1960_Girardeau_JMP_1}.  The relation between the two is given by
\begin{equation}\label{eq:BEwave_TG}
 \lim_{c \to \infty} \psi\left({\boldsymbol{x}}|{\boldsymbol{\lambda}}\right) = \prod_{i<j=1}^N \text{sgn}(x_i -x_j)   \frac{\det_{i,j=1}^N \left(e^{i x_i \lambda_j} \right)}{\sqrt{N!}} \epp
\end{equation}

A generic quench protocol consists in preparing the system in some initial state $| 0 \rangle $ which is not an eigenstate of \eqref{eq:LL_Ham}. At $t = 0$, the system is let unitarly evolve with the Hamiltonian  \eqref{eq:LL_Ham}. The eigenstate basis $|\boldsymbol{I}\rangle$  allows to compute the exact time evolution of a generic operator $\mathcal{O}$ as given by
\begin{equation}\label{eq:Double_sum}
\langle 0(t) | \mathcal{O} | 0(t) \rangle =
   \sum\nolimits_{{\boldsymbol{I}},{\boldsymbol{I}}' }    e^{- S_{\boldsymbol{I}}^\ast - S_{{\boldsymbol{I}}'} }   e^{i ( \omega_{{\boldsymbol{I}}}  - \omega_{{\boldsymbol{I}}'}) t}  \langle {\boldsymbol{I}} | \mathcal{O} | {\boldsymbol{I}}' \rangle \epc
\end{equation}
where we introduced the logarithm of the overlap coefficient $S_{\boldsymbol{I}} = -\log{\langle {\boldsymbol{I}} | 0\rangle}$ between a normalized Bethe state and the initial state and the time evolved initial state $| 0(t) \rangle = e^{- i H_{LL} t}| 0 \rangle $.
{The double summation over the whole Hilbert space is impossible to perform but we can get considerable simplifications by going to the thermodynamic limit} $L \to \infty$
with fixed density $N/L= D $ (we denote such a limit with $\lim_{\text{th}}$). In this limit the finite size Bethe states are replaced by their thermodynamic equivalents $| \rho \rangle$ specified by two distributions of rapidities $\rho(\lambda), \rho^h(\lambda)$ related to each other by the thermodynamic Bethe equations
\begin{equation}
\rho(\lambda)+\rho^h(\lambda)= \frac{1}{2 \pi} \left( 1 + \int_{-\infty}^{\infty} d\mu \frac{2 c \; \rho(\mu) }{(\lambda - \mu)^2 + c^2}  \right) \epc
\end{equation}
and normalized by the density of the gas
\begin{equation}
\int_{-\infty}^{\infty} d\mu \rho(\mu) = D \epc
\end{equation}
where we choose $D=1$ here for simplicity. 
In terms of finite size Bethe states $| \boldsymbol{\lambda} \rangle$ at finite size $N$ there is an exponential number $\sim \exp(S_{YY}[\rho])$ of them which are representative of the same thermodynamic state  $| \boldsymbol{\lambda} \rangle \to | \rho \rangle$. The extensive functional $S_{YY}$ is the Yang-Yang entropy given by
\begin{equation}
S_{YY}[\rho]= L \int_{-\infty}^{\infty} d\lambda \big( (\rho +  \rho^h) \ln (\rho + \rho^h) -   \rho \ln \rho
  - \rho^h \ln \rho^h   \big) \epp
\end{equation} 
With such a thermodynamic expression for the eigenstates we can perform the same steps as in  \cite{2013_Caux_PRL_110,2014_DeNardis_PRA_89} to obtain a  computationally much less expensive formula for the whole time evolution (only valid for simple \textit{weak} operators $\mathcal{O}$ \cite{2013_Caux_PRL_110})
\begin{align} \label{eq:QA_expectation}
&\langle 0(t) | \mathcal{O} | 0(t) \rangle = \frac{1}{2} \int \Drho \:   e^{ -2S[\rho] + \SYY[\rho]} \times\notag\\ 
&\sum_{ \mathbf{e} } \Big(   e^{ - \delta s_\mathbf{e} -  i \delta \omega_\mathbf{e} t } \langle \rho | \mathcal{O} | \rho , \mathbf{e} \rangle +   e^{ - \delta s^*_\mathbf{e} +  i \delta \omega_\mathbf{e} t } \langle \rho,\mathbf{e} | \mathcal{O} | \rho  \rangle \Big) \epc 
\end{align}
where $S[\rho]  =   \limth \Re S_I $ is the extensive real part of the overlap coefficient and $\mathbf{e}$ denotes a class of discrete excitations on the state $| \rho \rangle$ with
energy $\delta \omega_\mathbf{e}$ and differential overlap $\delta s_\mathbf{e} = - \log \big(  \langle  I \cup \mathbf{e} | 0 \rangle / \langle I | 0 \rangle \big)  $. 
In the thermodynamic limit the functional integral can be evaluated in its saddle point defined as $\left.\frac{\delta S^Q[\rho] }{\delta\rho} \right|_{\rhosp} = \left.\frac{\delta ( 2S[\rho] - \SYY[\rho])}{ \delta \rho }\right|_{\rhosp} = 0$ analogously to \cite{2013_Caux_PRL_110,2014_DeNardis_PRA_89} leading to a final expression for the whole time evolution after the quench 
\begin{equation}\label{eq:QA_expectation_2}
\limth\langle 0(t) | \mathcal{O} | 0(t) \rangle=\frac{1}{2}
\sum_{ \mathbf{e} } \Big(   e^{ - \delta s_\mathbf{e} -  i \delta \omega_\mathbf{e} t } \langle \rho_{sp} | \mathcal{O} | \rho_{sp} , \mathbf{e} \rangle +   e^{ - \delta s^*_\mathbf{e} +  i \delta \omega_\mathbf{e} t } \langle \rho_{sp},\mathbf{e} | \mathcal{O} | \rho_{sp}  \rangle \Big) \epp
\end{equation}
Expression \eqref{eq:QA_expectation_2} is exact in the thermodynamic limit and valid for any time $t > 0$ after the quench. \\

We focus now on the time evolution of the expectation values of some physical operators when the initial state $|0 \rangle $ is the bosonic ground state in the absence of interactions, {\it i.\,e.}~the BEC state 
\begin{equation}
|0 \rangle = |\text{BEC}\rangle \epc
\end{equation}
with $\langle \mathbf{x} | \text{BEC} \rangle = \frac{1}{L^{N/2}}$.
The limit $t \to \infty$ of expression \eqref{eq:QA_expectation_2} and the characterization of the saddle point state is given in \cite{2014_DeNardis_PRA_89} while here we focus on the time evolution towards the saddle point state.

The necessary excitations to resolve the whole time evolution can be written in terms of particle-hole excitations $\{ \mu^+_j, \mu_j^-\}_{j=1}^n$ where $n$ is a sub-extensive number $n \ll N $ such that 
\begin{equation}
\lim_{\text{th}} \frac{n}{N} = 0 \epp
\end{equation}
Given one of the many finite size $N=L$ normalized realizations of the saddle point state  $| \boldsymbol{\lambda}_{sp} \rangle \to | \rho_{sp} \rangle$ we define as holes a set of $n$ rapidities $\{ \lambda_{c_j} \}_{j=1}^n \equiv \{ \mu^-_j \}_{j=1}^n $ where $\{ c_j \}$ is a $n-$partition of $[1, \ldots
, N]$. We define then as particles the new values $ \{ \mu_j^+ \}_{j=1}^n$ that we assign to them $\{ \lambda_{c_j} \}_{j=1}^n \to \{ \mu_j^+ \}_{j=1}^n$. We denote such an excited state of $| \boldsymbol{\lambda}_{sp} \rangle$ with the notation  $| \boldsymbol{\lambda}'_{sp} , \{ \mu_j^-  \to \mu_j^+ \}_{j=1}^n \rangle$. The bulk rapidities which have not been chosen as holes $\boldsymbol{\lambda}'_{sp}$ get shifted by a $1/L$ factor according to 
\begin{equation}
\lambda'_{sp,j} - \lambda_{sp,j} = -\frac{1}{L}\sum_{k=1}^n \frac{F(\lambda_{sp,j}| \mu_k^+) -F(\lambda_{sp,j}| \mu_k^-) }{ \rho_{sp}(\lambda_{sp,j}) + \rho^h(\lambda_{sp,j})}  + \mathcal{O}(1/L^2) \epc
\end{equation}
where the back-flow $F(\lambda|\mu)$ is given in terms of the distribution $\rho(\lambda)$ of the thermodynamic state as \cite{KorepinBOOK}
\begin{equation}
2 \pi F(\lambda|\mu) =  \theta(\lambda - \mu) + \int_{-\infty}^{\infty} d\alpha \frac{2 c}{(\lambda  - \alpha)^2 + c^2} \frac{\rho(\alpha)}{\rho(\alpha)+ \rho^h(\alpha)}  F(\alpha|\mu)   \epp
\end{equation}

Since the overlaps of the BEC state with the Bethe states are non-zero only for parity-invariant Bethe states \cite{2014_DeNardis_PRA_89,1742-5468-2014-5-P05006,2014_Brockmann_npi} the only allowed type of particle-hole excitations to be included in the sum \eqref{eq:QA_expectation_2} are the parity-invariant ones
$\{ \mu_j^+ , - \mu_j^+, \mu_j^-, -\mu_j^- \}_{j=1}^n$. The differential overlaps and the energies of these type of excitations for a generic post-quench interaction strength $c$ are factorized (see Appendix A in \cite{2014_DeNardis_PRA_89})
\begin{align}\label{eq:deltasdeltae}
 &  e^{- \delta s_{\mathbf{e}} } =  \prod_{k=1}^n \exp \left( - \delta s(\mu_k^+) + \delta(\mu_k^-) \right)\epc \\
&
e^{- i \delta \omega_{\mathbf{e}} t}  = \prod_{k=1}^n  \exp\left( -2i t ( \delta \omega(\mu_k^+) - \delta \omega(\mu_k^-)  ) \right) \epc
\end{align}
where the differential overlap is given in terms of the distribution $\rho = \rho_{sp}$ by  
\begin{equation}
\delta s(\mu)= \frac{1}{2}\int_{-\infty}^\infty d\lambda \: \rho(\lambda)  \frac{1 + 8 \frac{\lambda^{2}}{c^{2}}}{\lambda \left(1+ 4\frac{\lambda^{2}}{c^{2}}\right)} F(\lambda | {\mu} ) 
+  \log \left({{\mu}  \sqrt{ ( {\mu}/c)^2 + 1/4}}\right)  \epp
\end{equation}
The differential energy $\delta \omega(\mu)$ is a functional of the distribution $\rho$ as given by the integral equation \cite{KorepinBOOK}
\begin{align}
&\delta \omega(\mu) = \mu^2 + 2\int_{-\infty}^{\infty} d\alpha  \: \alpha \: F(\alpha | \mu) \frac{\rho(\alpha)}{\rho(\alpha)+ \rho^h(\alpha)}     \epp 
\end{align}

The time evolution \eqref{eq:QA_expectation_2} is then written in terms of these excitations as 
\begin{align}\label{eq:laststep}
& \langle 0(t) | \mathcal{O} | 0(t) \rangle \nonumber 
\\&
=   \frac{1}{2}  \sum_{n=0}^\infty  \left[ \sum_{0<\mu_1^+ < \mu_2^+ < \ldots < \mu_n^+}  \sum_{0<\mu_1^- < \mu_2^- < \ldots < \mu_n^-}  \right]  
\langle \boldsymbol{\lambda}_{sp} | \mathcal{O} | \boldsymbol{\lambda}'_{sp} , \{ \mu_j^- , - \mu_j^- \to \mu_j^+, -\mu_j^+ \}_{j=1}^n \rangle
    \nonumber  + \text{mirr.}  \\&
=  \frac{1}{2}\sum_{n=0}^\infty \frac{1}{n!^2}  \left[\prod_{j=1}^n\sum_{\mu_j^- > 0} \sum_{\mu_j^+ > 0 } \right] 
\langle \boldsymbol{\lambda}_{sp} | \mathcal{O} | \boldsymbol{\lambda}'_{sp} , \{ \mu_j^- , - \mu_j^- \to \mu_j^+, -\mu_j^+ \}_{j=1}^n \rangle
e^{- \delta s_{\mathbf{e}} - i \delta \omega_{\mathbf{e}} t}   + \text{mirr.} \epc  
\end{align}
where $ \delta s_{\mathbf{e}}$ and $\delta \omega_{\mathbf{e}}$ are factorized for each excitation as in equations \eqref{eq:deltasdeltae} (which is believed to be valid for any type of quench to the Lieb-Liniger model). 
The mirrored sum is obtained by summing over the excitations on the left state instead of on the right state
\begin{equation}
 \text{mirr.}  = \frac{1}{2}\sum_{n=0}^\infty \frac{1}{n!^2}  \left[\prod_{j=1}^n\sum_{\mu_j^- > 0} \sum_{\mu_j^+ > 0 } \right] 
\langle \boldsymbol{\lambda}'_{sp}, \{ \mu_j^- , - \mu_j^- \to \mu_j^+, -\mu_j^+ \}_{j=1}^n  | \mathcal{O} | \boldsymbol{\lambda}_{sp} \rangle
e^{- \delta s_{\mathbf{e}} + i \delta \omega_{\mathbf{e}} t}  \epp  
\end{equation}
One should note that in the last step of equation \eqref{eq:laststep} we included in the sums the points $\mu_j^+ = \mu_k^+$ and  $\mu_j^- = \mu_k^-$ for any $j,k$. However these contributions are zero due to the fact that the matrix element when one of the states has two coinciding rapidites is zero \cite{KorepinBOOK}.
Assuming that the saddle point distribution is a smooth function (typical for any initial state with a non-zero energy density) the sums over the rapidities of the excitations can be converted to integrals for large system size leading to 
\begin{equation}
\left[\prod_{j=1}^n\sum_{\mu_j^- > 0} \sum_{\mu_j^+ > 0 } \right] \to L^{2n} \left[\prod_{j=1}^n \int_0^\infty d\mu_j^- d\mu_j^+ \rho_{sp}(\mu_j^-) \rho^h_{sp}(\mu_j^+)\right] \epp
\end{equation}
The system size divergence $L^n$ coming from the sum is reabsorbed into the form factors which have the same scaling in system size. The essential ingredients are then the form factors of physical operators in the thermodynamic limit for a given saddle point state and a given number $n$ of particle-hole excitations
\begin{align}\label{eq:therm_FF}
&\lim_{th} L^{n}\langle \boldsymbol{\lambda}_{sp} | \mathcal{O} | \boldsymbol{\lambda}'_{sp} , \{ \mu_j^- ,\to \mu_j^+ \}_{j=1}^n \rangle= \langle {\rho}_{sp} | \mathcal{O} | {\rho}_{sp}, \{ \mu_j^- \to\mu_j^+ \}_{j=1}^n \rangle \epc \\&
\lim_{th} L^{n} \langle \boldsymbol{\lambda}'_{sp} ,  \{ \mu_j^- , - \mu_j^- \to \mu_j^+, -\mu_j^+ \}_{j=1}^n  | \mathcal{O} | \boldsymbol{\lambda}_{sp} \rangle = \langle {\rho}_{sp},  \{ \mu_j^- \to\mu_j^+ \}_{j=1}^n | \mathcal{O} | {\rho}_{sp} \rangle \epp
\end{align}
With this notation we can finally express the time evolution after the quench in the thermodynamic limit
\begin{align}\label{eq:TEF}
&\lim_{\text{th}} \langle 0(t) | \mathcal{O} | 0(t) \rangle \nonumber \\&
=  \frac{1}{2}\sum_{n=0}^\infty \frac{1}{n!^2}  \left[\prod_{j=1}^n \int_0^\infty d\mu_j^+ \int_0^\infty d\mu_j^- \rho_{sp}(\mu_j^-) \rho_{sp}^h(\mu_j^+) e^{- \delta s(\mu_j^+) + \delta s(\mu_j^-) - 2 i t (\delta \omega(\mu_j^+) -\delta \omega(\mu_j^-)  )} \right] \nonumber
\\&
\times\langle {\rho}_{sp} | \mathcal{O} | {\rho}_{sp}, \{ \mu_j^- , - \mu_j^-\to\mu_j^+ , -\mu_j^+ \}_{j=1}^n \rangle  + \text{mirr.} \epc
\end{align}

Formula \eqref{eq:TEF} is valid in general for quenches in the Lieb-Liniger model with any interaction $c$ and any parity-invariant initial state. The computation of the form factors \eqref{eq:therm_FF} for two-point operators is however rather involved and up to now there is no generic method to obtain them. \\

From now on we focus on the quench from the BEC state to the $c=+\infty$ Lieb-Liniger model. In this limit the Bethe Ansatz wave function \eqref{eq:EF} takes the simpler form given in equation \eqref{eq:BEwave_TG} and it can be integrated  together with some static physical operators for any finite size $N$ (see \ref{Finite_Size} and Appendix C in \cite{2014_DeNardis_PRA_89}) \cite{zvonarev}. It is then possible to extract the thermodynamic limit of the finite size expressions for the form factors. Also in this limit the back-flow is zero $\lim_{c \to \infty }F(\lambda | \mu)=0$ and as a first consequence the differential overlaps and the energies  have much simpler expressions in terms of the rapidities of the excitations:
\begin{align}\label{eqn:BEC_Overlap}
&\exp \delta s(\mu)= \exp(\log \mu/2 ) = \mu/2 \epc \\&
 \exp (- 2 i t \delta \omega(\mu))  =\exp( -2 i t \mu^2) \epp 
\end{align}
The saddle point distribution for this quench is given by \cite{2013_Kormos_PRB_88,2014_DeNardis_PRA_89,2014_Kormos_PRA_89}
\begin{align}
&\rho_{sp}(\lambda) = \frac{1}{2 \pi} \frac{1}{1 + (\lambda/2)^2}  \epc \\&
\rho_{sp}^h(\lambda) =\frac{1}{2 \pi} \left(1 - \rho_{sp}(\lambda)  \right) = \frac{1}{2 \pi} \frac{(\lambda/2)^2}{1 + (\lambda/2)^2}  \epp
\end{align}

\section{Time Evolution of the density-density correlations} \label{sec3}
We review here some results obtained in \cite{2014_DeNardis_PRA_89} for the two-point density-density operator $\mathcal{O} =  \hat{\rho}(x)\hat{\rho}(0) $ where $\hat{\rho}(x) = \Psi^+(x)\Psi(x)$ and $\Psi$ is the bosonic annihilation operator. 

At $c= \infty$ the density operator $\hat{\rho}$ connects only states which differ at most for one particle-hole. The only non-zero form factors of $\hat{\rho}(x)\hat{\rho}(0)$ are then the diagonal one and the two particle-hole ones corresponding to the first two terms in the sum   \eqref{eq:TEF}. The form factor for the parity-invariant two-particle hole excitations is given by
\begin{equation}
  \langle \rho_{sp} |\hat{\rho}(x)\hat{\rho}(0)  | \rho_{sp} , \{ \mu^- , - \mu^- \to  \mu^+, -\mu^+ \} \rangle = {4}{} \sin{(\mu^- x)}\sin{(\mu^+ x)} \epc
\end{equation}
which is a special case of the two particle-hole form factor for generic thermodynamic states $| \rho \rangle$ at $c =\infty$ 
\begin{equation}
  \langle \rho |\hat{\rho}(x)\hat{\rho}(0)  | \rho ,\{ \mu_1^- ,  \mu_2^- \to \mu_1^+, \mu_2^+ \} \rangle = \Big(   e^{i\, x \mu^+_{1} } -    e^{i\, x \mu^+_{2} } \Big)\Big(  e^{-i\, x{\mu}^-_{1}} -   e^{-i\, x {\mu}^-_{2} } \Big) \epp
\end{equation}
 
Combining this with the diagonal form factor
\begin{equation}
\langle \rho|\hat{\rho}(x)\hat{\rho}(0)  | \rho \rangle =  1 + \delta(x) - \Big| \int_{-\infty}^\infty  d\lambda \rho(\lambda) e^{i  x \lambda} \Big|^2 \epc
\end{equation}
we can recover the whole time evolution summing over all these classes of possible excitations
\begin{align}
&\langle 0(t) | \hat{\rho}(x)\hat{\rho}(0)  | 0(t) \rangle= \langle \rho_{sp} |\hat{\rho}(x)\hat{\rho}(0)  | \rho_{sp} \rangle \\&+  4\int_{0}^\infty {d\mu^-}  {d\mu^+}  \rho_{sp}(\mu^-) \rho^h_{sp}(\mu^+)   e^{- 2it \epsilon(\mu^+)  + 2it \epsilon (\mu^-) - \delta s (\mu^+) + \delta s (\mu^-)  } \sin(\mu^+ x) \sin(\mu^- x)  \epc
\end{align}
where we used the fact that $\hat{\rho}(x)\hat{\rho}(0)$ is a self-adjoint operator to write the time evolution only as a single sum. 
By substituting the expressions for the BEC quench overlaps \eqref{eqn:BEC_Overlap} we can then simplify the above expression to
\begin{align}
&\langle 0(t) |\hat{\rho}(x)\hat{\rho}(0)  | 0(t) \rangle=  \delta(x) +  1- e^{- 4|x|}  + \frac{1}{4}\Big|e^{2 x} \text{erfc} \left( \frac{8 i t +x}{\sqrt{8 i t}} \right) - (x \to - x) \Big|^2 \epp
\end{align}
This reproduces the result of Ref.~\cite{2014_Kormos_PRA_89} but differently from the method used in there. The result can be extended easily to a generic quench protocol with a given saddle point distribution $\rho_{sp}$ and differential overlaps $\delta s$. 
It is interesting to consider the large time behaviour of the time evolution. The approach of the correlator to its saddle point value is indeed given by
\begin{align}\label{eq:large_t_density}
&\langle 0(t) |\hat{\rho}(x)\hat{\rho}(0)  | 0(t) \rangle - \langle \rho_{sp} | \hat{\rho}(x)\hat{\rho}(0)  | \rho_{sp} \rangle =    \frac{1}{512 \pi} \frac{1}{t} \left( \frac{x}{t} \right)^2 + \mathcal{O}(t^{-5})\epp
\end{align}

\begin{figure}[h!]
\includegraphics[scale=1]{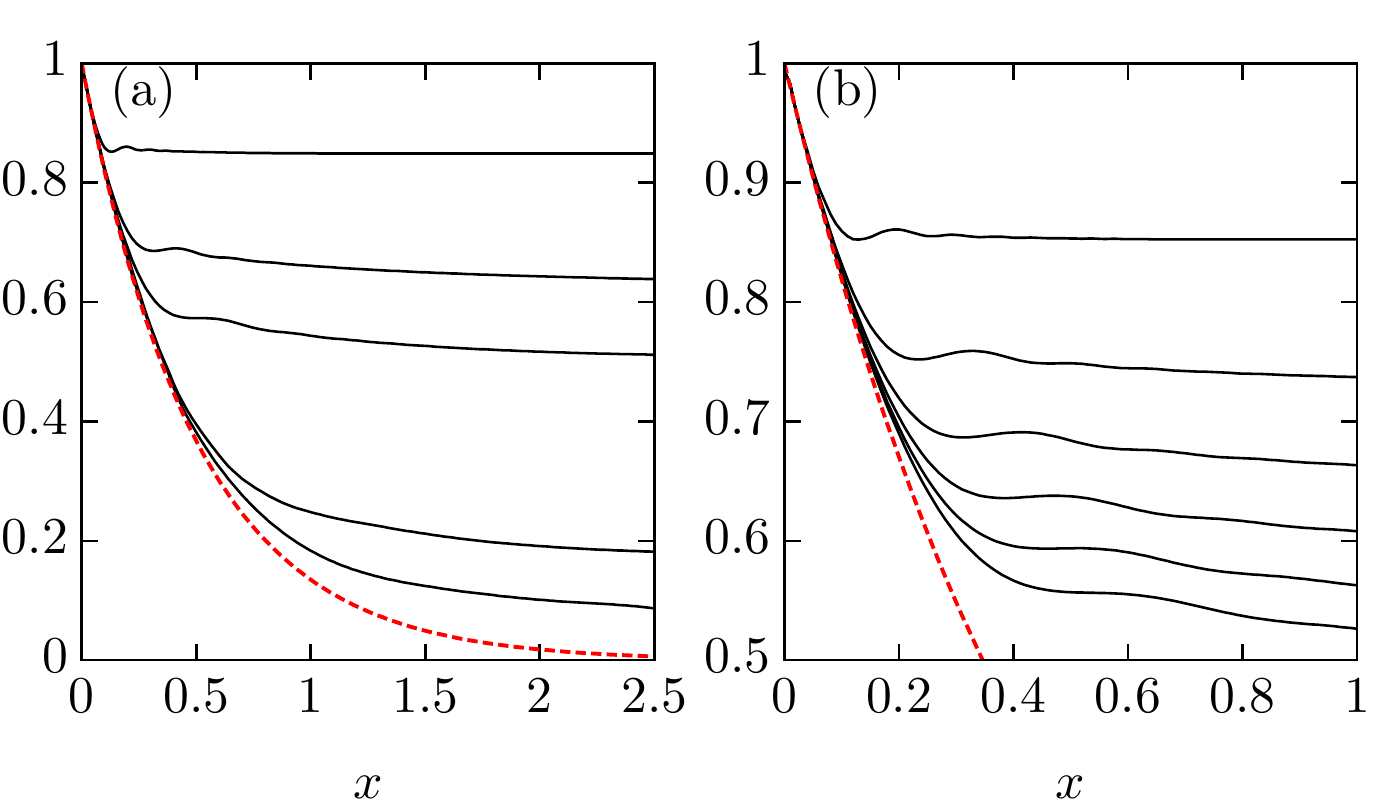}\caption{\label{fig:time_ev1} Panel (a): Time evolution of one-body density matrix $ \langle 0 (t)|   \Psi^+(x) \Psi(0)  | 0 (t) \rangle $ on the time evolved BEC state $\e^{-i t H_{LL}} |0 \rangle = | 0 (t) \rangle$ as a function of $x$ and for given values of time $t=10^{-3}, 5 \times 10^{-3}, 10^{-2}, 5 \times 10^{-2}, 10^{-1}$ (black lines)  from top ($t=10^{-3}$) to bottom $(t=10^{-1})$ and for infinite time after the quench $ \lim_{t \to \infty }\langle 0 (t)|   \Psi^+(x) \Psi(0)  | 0 (t) \rangle = e^{-2 |x|} $ (red dotted line).  Panel (b):  $ \langle 0 (t)|   \Psi^+(x) \Psi(0)  | 0 (t) \rangle$ on the interval $x\in[0,1]$ and for $t=10^{-3}( 1 + 2k) $ for $k \in [0,5]$  (black lines)  from top ($k=0$) to bottom $(k=5)$.}
\end{figure}


\section{Time evolution of the one-body density matrix}\label{sec4}

In the following we derive an analytical expression for the time evolution of the bosonic field-field operator $\mathcal{O} = \Psi^+(x) \Psi(y) $ (where we can set $y=0$ due to the translational invariance of the initial state) at $c=\infty$ which constitutes the main result of this paper. Differently from the density operator, $\Psi^+$ and $\Psi$ connect states differing by an arbitrary number of particle-hole excitations. The computation of the full time evolution then requires to perform the whole sum in \eqref{eq:TEF}. 

The Fourier transform of the one-body density matrix gives the single particle bosonic momentum distribution function which in general is not the same as the fermionic one and whose exact expression even for the ground state  or for  a generic thermal state is rather involved \cite{1964_Lenard_JMP_5,KorepinBOOK}. 

The necessary thermodynamic form factors are computed in \ref{FF_Psi}, here we report their expression for the parity-invariant excitations over the saddle point state $|\rho_{sp} \rangle $ : 
\begin{align}\label{eq:FF_psi_parity}
& \langle \rho_{sp} | \Psi^+(x) \Psi(0) | \rho_{sp} , \{ -\mu_i^-, \mu_i^- \to -\mu_i^+, \mu_i^+ \}_{i=1}^n \rangle    \\& = \text{Det} ( 1+ {K'} \rho_{sp})  \det_{i,j=1}^n  \begin{pmatrix}
W'(\mu^-_i,\mu^+_j) & W'(\mu^-_i,-\mu^+_j) \\ W'(-\mu^-_i, \mu^+_j) & W'(-\mu^-_i, -\mu^+_j) 
\end{pmatrix} \\&
- \text{Det} ( 1+ {K}\rho_{sp})  \det_{i,j=1}^n  \begin{pmatrix}
W(\mu^-_i,\mu^+_j) & W(\mu^-_i,-\mu^+_j) \\ W(-\mu^-_i, \mu^+_j) & W(-\mu^-_i, -\mu^+_j) 
\end{pmatrix}  \epc
\end{align}
where $K$ and $K'$ are two kernels given respectively by
\begin{align}
&K(u,v) = - 4 \frac{\sin \frac{x}{2} (u-v)}{u-v} \epc \\
& K'(u,v) = - 4 \frac{\sin \frac{x}{2} (u-v)}{u-v}  + e^{i \frac{x}{2} (u+v)} \epp
\end{align} 
The kernels $W= (1 + K \rho_{sp})^{-1} K$ and $W'= (1 + K' \rho_{sp})^{-1} K'$ are rigorously defined as solution of the following integral equations
\begin{align}
& W(u,v) + \int_{-\infty}^{\infty} dz \:  K(u,z) \rho(z) W(z,v) = K(u,v) \epc \\ 
& W'(u,v) + \int_{-\infty}^{\infty} dz \: K'(u,z) \rho(z) W'(z,v) = K'(u,v)  \epp
\end{align}
Given a kernel $A(u,v)$ and a function $\phi(u)$ we denote with $\text{Det}( 1+ A \phi)$ the Fredholm determinant (see \ref{Fredholm} and \cite{2008_Bornemann}) of the kernel $[A \phi](u,v) = A(u,v) \phi(v)$.  We also introduced the short-hand notation for the minor of a kernel  $A(u,v)$
\begin{equation}
\det_{i,j=1}^n A(x_i,x_j) = \det \begin{pmatrix}
A(x_1,x_1) & A(x_1,x_2) & \ldots & A(x_1, x_n) \\
\vdots  &  & & \vdots \\
A(x_n,x_1) & A(x_n,x_2) & \ldots & A(x_n, x_n) 
\end{pmatrix} \epp
\end{equation}

We can then perform the sum of the first term in \eqref{eq:TEF} and finally add the mirrored part. The property of the form factors under exchange of left and right state
\begin{equation}
\langle \rho , \{   \mu_i^- \to   \mu_i^+ \}_{i=1}^n | \Psi^+(x) \Psi(0) | \rho \rangle = \Big( \langle \rho| \Psi^+(x) \Psi(0) | \rho  , \{   \mu_i^- \to \mu_i^+ \}_{i=1}^n \rangle^* \Big)\Big|_{x \to - x}     \epc
\end{equation}
allows to express the mirrored sum as the complex conjugate with $x \to - x$ of the first sum.  Since this is a real and symmetric function of $x$ we can just express the time evolution of the one-body density matrix as a single sum
\begin{align}\label{eq:TE1}
&  \langle 0(t) |   \Psi^+(x) \Psi(0)  | 0(t) \rangle =\nonumber \\
= & \sum_{n=0}^\infty \frac{1}{n!^2}  \left[\prod_{j=1}^n \int_0^\infty d\mu_j^+ \int_0^\infty d\mu_j^- \rho_{sp}(\mu_j^-) \rho_{sp}^h(\mu_j^+) e^{- \delta s(\mu_j^+) + \delta s(\mu_j^-) - 2 i t (\delta \omega(\mu_j^+) -\delta \omega(\mu_j^-)  )} \right] \nonumber
\\&
\times\langle {\rho}_{sp} | \Psi^+(x) \Psi(0)  | {\rho}_{sp}, \{ \mu_j^- , - \mu_j^-\to\mu_j^+ , -\mu_j^+ \}_{j=1}^n \rangle  
\nonumber\\
&
= \text{Det}( 1+ K'\rho_{sp}) \sum_{n=0}^{\infty} \frac{1}{n!^2} \Big[\prod_{j=1}^n \int_{0}^{+ \infty}  d\mu_j^+ \rho_{sp}^h(\mu^+_j)    d\mu_j^- \rho_{sp}(\mu^-_j)     e^{-2i t (\delta \omega(\mu_j^+) -\delta  \omega (\mu_j^-)) - \delta s(\mu_j^+) + \delta s(\mu_j^-) } \Big]
\nonumber\\& \times 
\det_{i,j=1}^{n} \begin{pmatrix}
W'(\mu^-_i,\mu^+_j) & W'(\mu^-_i,-\mu^+_j) \\ W'(-\mu^-_i, \mu^+_j) & W'(-\mu^-_i, -\mu^+_j) 
\end{pmatrix}
-(K',W' \to K,W)  \epp
\end{align}
The product of the saddle point distribution times the  differential overlaps and energies can be rewritten as one function for particle excitations and one for holes
\begin{align}
& \rho_{sp}^h(\mu )     e^{-2i t \delta \omega(\mu) - \delta s (\mu)  }= e^{-2i t  \delta \omega(\mu) } \varphi_{+}^{(0)}(\mu) = \varphi_{+}^{(t)}(\mu) \epc \\&
 \rho_{sp}(\mu )     e^{2i t  \delta \omega(\mu) + \delta s (\mu)  }= e^{2i t  \delta \omega(\mu) } \varphi_{-}^{(0)}(\mu) = \varphi_{-}^{(t)}(\mu) \epp
\end{align}

Now we use the following identity \cite{MethaBOOK}: consider the following multi dimensional integral where $\mu(y)$ is some well-defined measure on a domain $X \in \mathbb{R}$
\begin{equation}\label{eq:ID_Pf}
\Big[ \prod_{\alpha=1}^n \int_X d\mu(y_\alpha) \Big] \det_{\alpha=[1,2n],\beta=[1,n]} \Big( A_\alpha(y_\beta)\: \: \: B_\alpha(y_\beta) \Big) \epp
\end{equation}
The integration can be performed and it leads to 
\begin{equation}
\Big[ \prod_{\alpha=1}^n \int_X d\mu(y_\alpha) \Big] \det_{\alpha=[1,2n],\beta=[1,n]} \Big( A_\alpha(y_\beta)\: \: \: B_\alpha(y_\beta) \Big) \epp= n! \sqrt{\det_{\alpha,\beta=1}^{2n} a_{\alpha \beta}}= n! \text{\text{Pf}}{}_{\alpha,\beta=1}^{2n} \: (a_{\alpha \beta}) \epc
\end{equation}
where $\text{\text{Pf}}{}_{i,j=1}^{2n} $ is the Pfaffian of the matrix $a_{\alpha \beta}$ and $a_{\alpha \beta}$ is given by
\begin{equation}
a_{\alpha \beta}= \int_X d\mu(y) \Big( A_\alpha(y)B_\beta(y) -    A_\beta(y)B_\alpha(y) \Big) \epp
\end{equation}
In our case we can choose $A_\alpha(y_\beta) = W(\mu_i^-, \mu^+_j )$($ W(-\mu_i^-, \mu^+_j )$) for odd(even) $\alpha \in [ 1, \ldots, 2n]$ and with the same logic $B_\alpha(y_\beta) = W(\mu^-_i, - \mu^+_j)$ ($B_\alpha(y_\beta) = W(-\mu_i^-,- \mu^+_j )$)  for odd(even) $\alpha \in [ 1, \ldots, 2n]$. With such a choice of the index we can integrate over the $\{ \mu_j^+\}_{j=1}^n$ for each $n$ using the identity \eqref{eq:ID_Pf}
\begin{align}\label{eqn:intermediate_1}
&\left[\prod_{j=1}^n \int_0^\infty {d\mu_j^{+}}{}  \varphi_{+}^{(t)}(\mu_j^+)\right] \det_{i,j=1}^{n} \begin{pmatrix}
W(\mu^-_i,\mu^+_j) & W(\mu^-_i,-\mu^+_j) \\ W(-\mu^-_i, \mu^+_j) & W(-\mu^-_i, - \mu^+_j) 
\end{pmatrix} \\
&= n! \:\: \text{Pf}_{i,j=1}^{n} \begin{pmatrix}
\Phi(\mu^-_i,\mu^-_j) & \Phi(\mu^-_i,-\mu^-_j) \\ \Phi(-\mu^-_i, \mu^-_j) & \Phi(-\mu^-_i, - \mu^-_j) 
\end{pmatrix} \epc
\end{align}
where the new kernel $\Phi(u,v)$ is given by
\begin{align}\label{eq:Phi_def}
 \Phi(u,v) &=  \int_0^\infty dy \:  \varphi_{+}^{(t)}(y)  \Big(  W(u,y) W(v,-y) -  W(u,-y) W(v,y) \Big) \\&
= \int_{-\infty}^\infty dy  \: \varphi_{+}^{(t)}(y)   W(u,y) W(v,-y)  \epp
\end{align}
In the last step we used the antisymmetry of the differential overlaps which leads to $  \varphi_{+}^{(t)}(-y) =-\varphi_{+}^{(t)}(y)$. The same can be done for the function $W'$ leading to an analogous result as in \eqref{eqn:intermediate_1} with a different kernel $\Phi'$ given by
 \begin{align}
 \Phi'(u,v) 
= \int_{-\infty}^\infty dy \:  \varphi_{+}^{(t)}(y)  W'(u,y) W'(v,-y) \epp
\end{align}
The expression \eqref{eq:TE1} then translates to
\begin{align}
& \langle 0(t) | \Psi^+(x) \Psi(0)  | 0(t) \rangle = \nonumber \\& 
= \text{Det}( 1+ K' \rho_{sp}) \sum_{n=0}^{\infty} \frac{1}{n!} \left[ \prod_{j=1}^n \int_{0}^{+ \infty}  d\mu_j^-    \varphi_{-}^{(t)}(\mu_j^-) \right]  
 \text{Pf}_{i,j=1}^{n} \begin{pmatrix}
\Phi'(\mu^-_i,\mu^-_j) & \Phi'(\mu^-_i,-\mu^-_j) \\ \Phi'(-\mu^-_i,  \mu^-_j) & \Phi'(-\mu^-_i, - \mu^-_j)  \nonumber
\end{pmatrix}\\&
  - (K',\Phi' \to K,\Phi)   \epp
\end{align}
We now use the definition of the Fredholm Pfaffian (see \ref{Fredholm}) for the kernel $\Phi$ (which can be equivalently implemented for the kernel $\Phi'$)  
\begin{align}\label{eq:stepsign}
&\sum_{n=0}^{\infty} \frac{1}{n!} \left[\prod_{j=1}^n \int_{0}^{+ \infty}  d\mu_j^-  \varphi_{-}^{(t)}(\mu_j^-) \right] 
\times \text{Pf}_{i,j=1}^{n} \begin{pmatrix}
\Phi(\mu^-_i,\mu^-_j) & \Phi(\mu^-_i,-\mu^-_j) \\ \Phi(-\mu^-_i,  \mu^-_j) & \Phi'(-\mu^-_i, - \mu^-_j) \nonumber
\end{pmatrix} \\&
= \text{Pf}\Big( \boldsymbol{J} +P_0 \boldsymbol{\Phi} \varphi_{-}^{(t)}  P_0\Big)  = \sqrt{\text{Det}(\boldsymbol{I} -  P_0 \boldsymbol{J}\boldsymbol{\Phi}  \varphi_{-}^{(t)}  P_0)} \epc
\end{align}
where we introduced the $2 \times 2$ matrix kernels
\begin{equation}
\boldsymbol{\Phi} = \begin{pmatrix}
 \Phi_{++} &   \Phi_{+-}  \\  \Phi_{-+}&  \Phi_{--}
\end{pmatrix}  \:\: \:\: \:\: \:\: \:\:  \boldsymbol{J} = \begin{pmatrix}
 0  & 1 \\ -1 & 0
\end{pmatrix} \epc
\end{equation}
where $1$ denotes the identity on the kernel space as usual and the set of kernels $\Phi_{\pm \pm}$ are defined as $\Phi_{\pm \pm} \equiv \Phi(\pm u, \pm v)$. The function $P_0$ is the projector on the positive real line $x>0$.
In the last step we used the relation between Fredholm Pfaffian and Fredholm determinant $ \text{Pf}\Big( \boldsymbol{J} +  \boldsymbol{\Phi}  \Big)^2  = {\text{Det}(\boldsymbol{I} -    \boldsymbol{J}\boldsymbol{\Phi}  )}$ as given in \ref{Fredholm}. The square root of this expression produces an undetermined sign which in \eqref{eq:stepsign} is chosen to be positive. This choice is purely arbitrary and its correctness is checked in the limit $t = 0^+$ where the expression for the time evolution recovers the one-body density matrix of a BEC state. The antisymmetry of $ \varphi_{-}^{(t)}(-y)= -   \varphi_{-}^{(t)}(y)$ leads to
\begin{equation}
\text{Det}(\boldsymbol{I} -P_0 \boldsymbol{J}\boldsymbol{\Phi}P_0 \varphi_{-}^{(t)} ) = \text{Det} \begin{pmatrix}
 1 - P_0   \varphi_{-}^{(t)}\Phi_{-+} P_0  &  - P_0  \varphi_{-}^{(t)} \Phi_{--} P_0  \\  P_0  \varphi_{-}^{(t)}  \Phi_{++} P_0  & 1 + P_0 \varphi_{-}^{(t)}  \Phi_{+-} P_0 
\end{pmatrix}  = \text{Det}( 1+ \varphi_{-}^{(t)}  \Phi_{+-} ) \epc
\end{equation}
where the last Fredholm determinant is defined on the whole $\mathbb{R}$ axis.
Using the behaviour of $W$ and $W'$ under inversion of its coordinates $W(-u,-v) = W^*(u,v)= (1 + K \rho_{sp})^{-1}{}^* K^*$ (see \ref{FF_Psi}) the kernel $\Phi_{+-} = \Phi(u,-v)$ defined in \eqref{eq:Phi_def} can be written in terms of the kernels $K$ as the following product of operators
\begin{equation}
\Phi_{+ -} = W \varphi_{+}^{(t)}  W^*   =  (1+K \rho_{sp})^{-1}{} K \varphi_{+}^{(t)}  [ (1+K \rho_{sp})^{*}]^{-1} K^* \epp
\end{equation}
Using the Cauchy-Binet formula for determinants we obtain
\begin{align}
&   \text{Det}( 1 + \varphi_{-}^{(t)} \Phi_{+-}  ) = \text{Det}( 1 + \Phi_{+-} \varphi_{-}^{(t)} ) 
=
\\& \frac{\text{Det}( 1 + K \rho_{sp} +  K \varphi_{+}^{(t)} [ (1+K \rho_{sp})^{*}]^{-1} K^*  \varphi_{-}^{(t)} ) }{\text{Det}( 1+K \rho_{sp})}  \nonumber
=  \frac{ \text{Det} \begin{pmatrix}
 1 + K\rho_{sp}  &  -  K \varphi_{+}^{(t)} \\   K^* \varphi_{-}^{(t)}  & 1  +  K^*\rho_{sp}
\end{pmatrix}  }{\text{Det}( 1+ K \rho_{sp})^2}  \epc
\end{align}
where in the last step we used the following property for the determinant of a block matrix 
\begin{equation}
\det \begin{pmatrix}
A & B \\ C & D
\end{pmatrix} = \det D \times \det ( A - B D^{-1} C) \epp
\end{equation}

\begin{figure}[h!]
\includegraphics[scale=1]{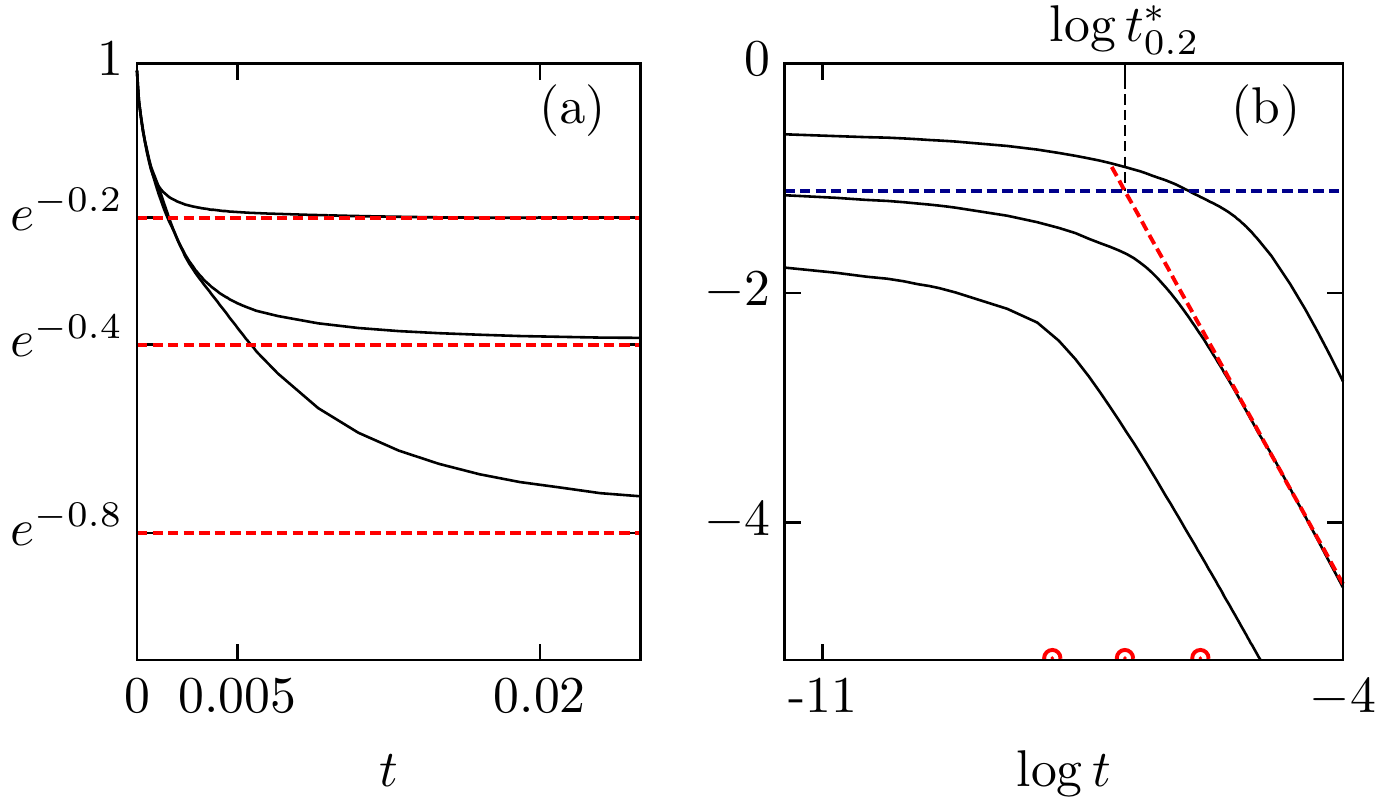}\caption{\label{fig:time_ev2_fig} Panel (a): Time evolution of the one-body density matrix  $ \langle 0 (t)|   \Psi^+(x) \Psi(0)  | 0 (t) \rangle $ on the time evolved BEC state $\e^{-i t H_{LL}} |0 \rangle = | 0 (t) \rangle$ as a function of time $t$ and for given values of $x=0.1,0.2,0.4$ (black lines)  from top ($x=0.1$) to bottom $(x=0.4)$. The red dotted lines show the respectively saddle point values. Panel (b): Time evolution of the logarithm of the time dependent correlation after having subtracted its saddle point value   $\log \left(\langle 0 (t)|   \Psi^+(x) \Psi(0)  | 0 (t) \rangle - e^{- 2 |x|} \right)$ as a function of $\log t$ and for given values of $x=0.4,0.2,0.1$ (black lines)  from top ($x=0.4$) to bottom $(x=0.1)$. For each value of the spatial separation $x$ there is a crossover time $t^*_x$ after that the correlation function approaches its saddle point value with the power law $t^{-7/6}$. The red dotted line shows the fitting function $ - 9.3 - 7/6 \log t$ while the value of the correlation in $x=0.2$ at $t=0$, given by $\log (1 - e^{-0.4})$, is shown by the blue dotted line. The logarithm of the crossover time $t^*_{0.2}$ is chosen to be the intersection point between the two lines and the same is done for the correlations at different points $x$. The obtained values of the logarithm of the crossover times $\log t^*_x$ are shown by the red semi-circles on the $\log t$ axis for $x=0.1,0.2,0.4$ (from left to right). The plot shows that the crossover times are linear functions of the spatial separation $x$ ($t^*_{2x} = 2 t^*_{x}$ up to the precision of the numerical evaluation) which is an indication of a light-cone spreading of information after the quench similarly to what is shown in\cite{2006_Calabrese_PRL_96,NATURE_397}}
\end{figure}

Then finally we can write the full analytical result for the time evolution valid at any time $t$ in the thermodynamic limit
\begin{align}\label{eq:TE_Final}
& \langle 0(t) |  \Psi^+(x) \Psi(0) | 0(t) \rangle \nonumber \\&
=  \sqrt{ \text{Det} \begin{pmatrix}
 1 + K' \rho_{sp}  &   - K' \varphi_{+}^{(t)} \\   K'^* \varphi_{-}^{(t)}   & 1  +   K'^* \rho_{sp}
\end{pmatrix}  } -\sqrt{  \text{Det} \begin{pmatrix}
 1 + K \rho_{sp}  & -   K  \varphi_{+}^{(t)}\\   K \varphi_{-}^{(t)}  & 1  +  K  \rho_{sp}
\end{pmatrix}  } \epp
\end{align}
Formula \eqref{eq:TE_Final} represents the main result of this paper. 
\\

 The limit $t \to \infty$ recovers the known results for the expectation value of the field-field operator on the saddle point state \cite{2013_Kormos_PRB_88,2014_DeNardis_PRA_89,2014_Kormos_PRA_89}
\begin{align}
& \lim_{t \to \infty}\langle 0(t) |   \Psi^+(x) \Psi(0)   | 0(t) \rangle \nonumber \\&
=  \sqrt{ \text{Det} \begin{pmatrix}
 1 + K' \rho_{sp}  &    0 \\  0  & 1  + K'^* \rho_{sp}
\end{pmatrix}  } -\sqrt{  \text{Det} \begin{pmatrix}
 1 + K \rho_{sp}  & 0\\   0 & 1  + K \rho_{sp}
\end{pmatrix}  } \nonumber \\&
= \text{Det}( 1+ K' \rho_{sp} ) -  \text{Det}( 1+ K \rho_{sp} )  \equiv \langle \rho_{sp} |  \Psi^+(x) \Psi(0)  | \rho_{sp} \rangle   = e^{- 2 |x|} \epc
\end{align}
where we set the kernels $K^*\varphi^{(t)}_+ $ and $K^*\varphi^{(t)}_- $ to zero since for any smooth function $g$ on $\mathbb{R}$ the oscillating phase $e^{-2 i t y^2}$ sets the action of the two kernels to zero for large $t$
\begin{equation}
 \lim_{t \to \infty} [K\varphi^{(t)}_\pm]  g =   \lim_{t \to \infty} \int_{-\infty}^{\infty} dz [K\varphi^{(t)}_\pm] (x,z) g(z) =0 \epc
\end{equation}
where the same is valid obviously also for the kernel $K'$. 

The first corrections to the steady state expectation for large time are in principle obtainable from the final formula \eqref{eq:TE_Final}. The numerical evaluations of it suggest that the saddle point expectation value is approached with corrections of order $\sim t^{-7/6}$ for large time. 


The limit $t \to 0^+$ is more involved to recover analytically but it can be evaluated numerically as for all other values of $t$. We discretized the kernels in \eqref{eq:TEF} with $m$ points on the $\mathbb{R}$ axis reducing in this way the Fredholm determinant to a  determinant of an $m \times m$ matrix as is explained in \cite{2008_Bornemann}. With a Gaussian quadrature method with $m = 800$ points in the range $x \in [-300,300]$ we obtain the expected one-body density matrix for a BEC state at $t=0$
\begin{equation}
\langle 0 |    \Psi^+(x) \Psi(0)   | 0 \rangle = 1 + \epsilon \epc
\end{equation}
with a constant numerical error $\epsilon \sim 10^{-3}$. 
From the same numerical procedure we obtain also the following analytical guesses for the Fredholm determinant of the two kernels (up to the numerical precision of the evaluation procedure)
\begin{align}
&\text{Det}( 1 + K \rho_{sp}) = e^{-2 |x|} \cos 2 x  \epc \\
&\text{Det}( 1 + K' \rho_{sp}) = e^{-2| x|}(\cos 2 x + 1) \epp
\end{align}
We used the same numerical procedure to obtain the plots in Fig \ref{fig:time_ev1} for different values of $t$ and $x$.

The limit $x \to 0$ gives the density of the gas $D$ and it is also easily recovered for any time $t$ since $K \to 0$ and $K' \to 1$. We then obtain
\begin{align}\label{densityxzero}
& \langle 0(t) |  \Psi^+(0) \Psi(0) | 0(t) \rangle \nonumber \\&
=  \sqrt{ \text{Det} \begin{pmatrix}
 1 +  \rho_{sp}  &   -  \varphi_{+}^{(t)} \\    \varphi_{-}^{(t)}   & 1  +   \rho_{sp}
\end{pmatrix}  } -1 \nonumber \\& = 
\sqrt{ \text{Det} \left[\begin{pmatrix} e^{- i t \delta \omega} & 0 \\ 0 & e^{ i t \delta \omega} \end{pmatrix} \begin{pmatrix}
 1 +  \rho_{sp}  &   -  \varphi_{+}^{(0)} \\    \varphi_{-}^{(0)}   & 1  +   \rho_{sp}
\end{pmatrix} \begin{pmatrix} e^{- i t \delta \omega} & 0 \\ 0 & e^{ i t \delta \omega} \end{pmatrix}^{-1}  \: \right] } -1 \nonumber
 \\& = 
\sqrt{ \text{Det}  \begin{pmatrix}
 1 +  \rho_{sp}  &   -  \varphi_{+}^{(0)} \\    \varphi_{-}^{(0)}   & 1  +   \rho_{sp}
\end{pmatrix}  } -1  = \sqrt{ \left(1 + \int_{-\infty}^\infty d\lambda \rho_{sp}(\lambda) \right)^2 + \left(\int_{-\infty}^\infty d\lambda \varphi_+^{(0)}(\lambda) \right)^2 } - 1 \nonumber \\
& = D  \epc
\end{align}
where in the last steps we used that $\varphi_{+}^{(0)}(\lambda) =\varphi_{-}^{(0)}(\lambda) $ and $\varphi_{+}^{(0)}(\lambda) = - \varphi_{+}^{(0)}(-\lambda)$. Furthermore we extended the following property of the determinant (with $\{ v_j \}_{j=1}^N$ and $\{ c_j \}_{j=1}^N$ two different vectors)
\begin{equation}
\det_{2 N} \begin{pmatrix} 
\delta_{ij} +  v_j  &   -  c_j\\    c_j  & \delta_{ij}  +  v_j
\end{pmatrix}  = \left( 1 + \sum_{j=1}^N v_j \right)^2 +\left( \sum_{j=1}^N c_j \right)^2 \epc
\end{equation}
to the Fredholm determinant in \eqref{densityxzero}.

\section{Conclusions}
In this paper we applied the quench action logic to compute the exact time evolution of the one-body density matrix for the interaction quench in the Lieb-Liniger model when the interaction strength $c$ is switched from $c=0$ to $c=+\infty$. Although the system is mappable to free fermions in this limit the one-body operator cannot be expressed easily in terms of the fermionic operators. Therefore this result constitutes the fist application of the quench action method to obtain a new exact result on the time evolution of operators which resisted to all the analytical approaches up to now. 
Moreover the methods used here are completely general and they can be used for any other quench protocol in the Tonks-Girardeau gas for hard-core bosons.

It would be interesting to compare this result with the one obtained by performing a corresponding quench in the interaction parameter in the Luttinger Liquid theory \cite{2009_Iucci_PRA_80,PhysRevB.88.115144,2012_Rentrop_NJP_14,2012_Karrasch_PRL_109}. The one-body bosonic function, which corresponds to the two-point correlation function of the phase fluctuations, is in principle directly computable in this framework and the exact result in this paper could then give some indications on the universality of time dependence for quenches in the Luttinger Liquid model. The numerical evaluations of the final formula in Fig. \ref{fig:time_ev2_fig} suggest that the phase-phase correlations spread after the quench according to a light-cone dynamical effect as is the case in an hydrodynamical theory \cite{NATURE_397}. This is in contrast with what has been observed for the time evolution of the density-density correlation functions within the same quench protocol \cite{2014_Kormos_PRA_89}. The dramatic difference between the two behaviors could be due to the simple structure of the density operator in the $c \to \infty$ limit. In this limit indeed the density operator becomes quasi-diagonal and it can only create a small number of excitations on the saddle point state after the quench (See Section \ref{sec3}), in contrast with the infinite number of them created by the field operator. Consequently the relaxation of the density-density correlations is much faster than the one of the one-body density function. 

 However for a more extensive analysis it is important to obtain expansions of the formula around the asymptotic points $t \to \infty$ and $x,t \to \infty$ with $x/t = v$ where universal features are expected and check the numerical evaluations of Fig. \ref{fig:time_ev2_fig}.  Due to the complicated dependence in time of the two Fredholm determinants we are not able to give a rigorous explanation of the power law scaling $t^{-7/6}$ for the approach to the steady state correlation function. We will come back to these questions in future publications.

Finally we will also address in future the extension of the quench action method to time evolution after quenches in the Lieb-Liniger model with finite coupling constant $c$.

\section*{Acknowledgements}
We acknowledge useful and inspiring discussions with Sebastiaan Eli\"{e}ns, Mi\l osz Panfil, Bram Wouters, Michael Brockmann, Davide Fioretto, Dirk Schuricht, Pasquale Calabrese. 

We acknowledge support from the Foundation for Fundamental Research on Matter (FOM) and the Netherlands Organisation for Scientific Research (NWO).

This work forms part of the activities of the Delta Institute for Theoretical Physics (D-ITP).

\appendix

\section{The field-field form factor in the Tonks-Girardeau regime}\label{FF_Psi}
The necessary step in order to compute the exact time evolution \eqref{eq:TE_Final} is the analytical expression of the thermodynamic limit of the form factor as defined in \eqref{eq:therm_FF} for the two bosonic fields at a distance $x$ for some representative finite size state $|  \boldsymbol{\lambda} \rangle  \to  | \rho \rangle $ of a generic thermodynamic state specified by a distribution $\rho(\lambda)$
\begin{equation}
\lim_{\text{th}} L^{n}\langle \boldsymbol{\lambda} | \Psi^+(x) \Psi(0) | \boldsymbol{\lambda},\{ \mu_j^- \to \mu_j^+ \}_{j=1}^n \rangle  \epp
\end{equation}
Given a finite system with $N$ particles (where $N$ is chosen to be even for simplicity) the expression for the form factor can obtained by integrating the Bethe wave function (See \ref{Finite_Size})
and it is given in terms of a difference of two determinants of two $N\times N$ matrices
\begin{equation}
\langle \{ \lambda_i\}_{i=1}^N | \Psi^+(x) \Psi(0) |  \{\mu_i \}_{i=1}^N \rangle = \det_N (V_1 + V_2) - \det_N V_1  \epc
\end{equation}
where the two matrices are given by
\begin{equation}
V_1  = - \frac{i \left(e^{-i L (\lambda_a-\mu_b  )}-2 e^{-i x (\lambda_a-\mu_b )}+1\right)}{L (\lambda_a-\mu_b )}  \epc
\end{equation}
\begin{equation}
V_2 = \frac{e^{ -i x \lambda_a}}{L} \epp
\end{equation}
The two sets $\{ \lambda_j\}_{j=1}^N$ and $\{ \mu_j\}_{j=1}^N$ are two different sets of parameters which do not satisfy Bethe equations in general. 

We can just slightly modify the two matrices by multiplying the determinants by $\prod_{a=1}^N e^{- i \frac{x}{2} (\lambda_a - \mu_a)}$ such that we have
\begin{equation}
\det_N  (V_1 +  V_2) - \det_N  V_1  = \prod_{a=1}^N e^{ i \frac{x}{2} (\mu_a - \lambda_a)} \left( \det_N \left( f(\lambda_a, \mu_b) +  p(\lambda_a) p(\mu_b) \right) -  \det_N  f(\lambda_a, \mu_b) \right) \epc
\end{equation}
where we introduced the notation
\begin{equation}
 f(\lambda, \mu) = - e^{- i \frac{x}{2} (\mu - \lambda)}  \frac{i \left(e^{-i L (\lambda-\mu  )}-2 e^{-i x (\lambda-\mu )}+1\right)}{L (\lambda-\mu )}  \epc
\end{equation}
\begin{equation}
p(\lambda) =\frac{1 }{L^{1/2}} e^{- i \frac{x}{2} \lambda}   \epp
\end{equation}

We now choose the set of rapidities of the left state as a representative state for a generic thermodynamic state $| \rho  \rangle$ specified by a distribution of rapidities $\rho(\lambda)$ and we choose the rapidities of the right state in a way to have a sub-extensive number of excitations $n\ll N$ compared to the left state. This corresponds to choosing $n$ rapidities from the set $\{ \lambda_a \}_{a=1}^N$ to be modified to some new value
\begin{align}
\mu_{c_i} = \mu_i^{+}  & \:\:\:\:\:\:
\lambda_{c_i} =  \mu_i^{-} &  i=1, \ldots, n \epc
\end{align}
and keeping the remaining  $N-n$ rapidities with the same value as in the left state
\begin{equation}
\mu_a = \lambda_a  \:\:\: \:\:\:\:\:\:\:\:\: a \neq \{ c_i \}_{i=1}^n \epp
\end{equation}
The possible choices $c_i$ for each $i$ take values in the whole set $[1,\ldots , N]$  with the condition  $c_i\neq c_j $ for each $i,j$.

Let us consider now the matrix $ f (\lambda_a , \mu_b)$. With such a choice of the right state we have that the matrix has different forms depending on which column we consider. Indeed for the columns $b$ corresponding to rapidities  $\mu_b$ in the right state that we do not excite we have
\begin{equation} \label{eq:excited_m1}
f(\lambda_a,\mu_b) = \delta_{ab} - \frac{4}{L} \frac{\sin(\frac{x}{2}(\lambda_a - \mu_b))}{(\lambda_a - \mu_b)}  \equiv \delta_{ab} + \frac{1}{L} K(\lambda_a, \lambda_b) \:\:\:\:\:\:\:\: b \neq c_j \epc
\end{equation}
while for the $n$ coloumns corresponding to the rapidities $\lambda_{c_i}$ we have
 \begin{equation} \label{eq:excited_m2}
f(\lambda_a,\mu_b) = \frac{1}{L} K(\lambda_a, \mu_j^+) \:\:\:\:\:\:\:\:  \:\:\:\:\:\:\:\: b = c_j \epc
\end{equation}
where eventually we used that all the rapidities satisfy Bethe equations
\begin{equation}
e^{i \lambda L} = -1 \epc
\end{equation}
and we introduced the kernel $K$ defined as
\begin{equation}
K(u,v) =-4 \frac{\sin(\frac{x}{2}(u-v)) }{u-v} \epp
\end{equation}
We can bring the factors $1/L$ in the columns $b=\{c_j\}_{j=1}^n$ outside the determinant. Then this can then be written as 
\begin{equation}\label{eq:det1}
\det  f (\lambda_a , \mu_b) = L^{-n} \det \begin{pmatrix}
 \delta_{ab} + \frac{K(\lambda_a, \lambda_b)}{L}   + B_{ab} 
 \end{pmatrix} \epc
\end{equation}
where we added and subtracted the matrix with no excitations $\delta_{ab} + \frac{1}{L} K(\lambda_a, \lambda_b)$. The matrix $B$ is then the difference between the latter and the matrix with excitation given in \eqref{eq:excited_m1} and \eqref{eq:excited_m2}. The two matrices differ only in the columns corresponding to the set of indices of the excited rapidities $\{c_j\}_{j=1}^n$. Therefore the matrix $B$ has only $n$ non-zero columns given by
\begin{align}
B_{a c_j}   & = K(\lambda_a, \mu_j^+) - \frac{1}{L}  K(\lambda_a, \lambda_{c_j}) \:\:\:  \:\:\: a \neq c_j \epc\\
B_{a c_j}  & = K(\lambda_{c_j} , \mu_j^+)  -1 - \frac{1}{L}  K(\lambda_{c_j} , \lambda_{c_j})  \:\:\:\:\:\:a = c_j \epc \\
B_{a b}  & = 0  \:\:\:\:\:\:  \:\:\:\:\:\: b \neq  c_j \epp
\end{align}
Using standard properties of the determinant we can decompose the determinant in \eqref{eq:det1} as a product of a determinant of a $N \times N$ matrix and the determinant of a $n \times n$ one as follows
\begin{align}
&\det_N  f (\lambda_a , \mu_b) = \frac{1}{L^n} \det_N \left( \delta_{ab } + \frac{1}{L} K(\lambda_a, \lambda_b) + B_{ab}  \right) \nonumber \\
= 
&\frac{1}{L^n} \det_N \left( \delta_{ab } + \frac{1}{L} K(\lambda_a, \lambda_b)  \right) \det_n  \left(\delta_{ij} +\sum_{l=1}^N \left(1+ \frac{K}{L} \right)^{-1}_{c_i,l} B_{lj} \right) \epc
\end{align}
and where $1$ denotes here the identity matrix. Since the number of excitations is sub-extensive, in the thermodynamic limit we can neglect all the $1/L$ corrections in the determinant of the $n\times n$ matrix. Therefore we obtain
\begin{equation}\label{eq:Wintro}
\lim_{\text{th}}\det_n \left(\delta_{ij} + \sum_{l=1}^N  \left(1+ \frac{K}{L} \right)^{-1}_{c_i,l}B_{lj} \right) = \det_{i,j=1}^n W(\mu_i^- , \mu_j^+) \epc
\end{equation}
and for the $N \times N$ matrix we obtain a Fredholm determinant
\begin{equation}
\lim_{\text{th}} \det_N \left( \delta_{ab } + \frac{1}{L} K(\lambda_a, \lambda_b)  \right) = \text{Det}( 1+ {K} \rho) \epp
\end{equation}
The function $W$ introduced in \eqref{eq:Wintro} satisfies the following integral equation
\begin{align}
W(u , v)&  + \int ds  K( u, s) \rho(s )W(s, v)  = K(u , v)  \epc
\end{align}
which can be equivalently rewritten in matrix formalism as 
\begin{equation}
W = (1 + K \rho)^{-1} K \epp
\end{equation}
The same procedure can be done for the determinant of the matrix $f(\lambda_a, \mu_b) + p(\lambda_a)p(\mu_b)$ just by shifting the kernel by the rank one kernel $e^{-\frac{x}{2} i (u + v)} $
\begin{equation}
 K' (u, v) = K(u, v) +e^{-\frac{x}{2} i (u + v)}  \epp
\end{equation}
We denote the corresponding $W$ kernel associated to $K'$ as $W' = (1 +K' \rho )^{-1} K' $. 
From the behaviour of the two kernels under a change of sign in their arguments 
\begin{align}
&K(-u,-v) = K(u,v) \epc \\
&K'(-u,-v) = K'(u,v)^* \epc
\end{align}
we obtain $W(-u,-v)  = W(u,v)$ and $W'(-u,-v)  = W'^*(u,v)$.  Since the two kernels are symmetric $K(u,v)=K(v,u) $, $K'(u,v)=K'(v,u) $ $W$ and $W'$ are symmetric as can be seen by a formal geometric series expression in $K$ (or $K'$) 
\begin{align}
& W = ( 1+ K \rho)^{-1}K = K - K \rho K + K \rho K \rho K + \ldots \epc \\
& W' = ( 1+ K' \rho)^{-1}K' = K' - K' \rho K' + K' \rho K' \rho K' + \ldots \epp
\end{align}

Finally the matrix element for $n$ particle-hole excitations on a generic thermodynamic state $| \rho \rangle$ of the two-point field-field operator is given by
\begin{align}
& \langle \rho  | \Psi^+(x) \Psi(0) | \rho , \{ \mu_i^-\to \mu_i^+\}_{i=1}^n \rangle =  e^{ i \frac{x}{2}\sum_{j=1}^n ( \mu_j^+ - \mu_j^-)} \nonumber \\& \times  
\Big[\text{Det} ( 1+ {K'} \rho)  \det_{i,j=1}^n [ W'(\mu_i^+, \mu_j^-)] - \text{Det} ( 1+ {K}\rho )  \det_{i,j=1}^n [ W(\mu_i^+, \mu_j^-)]\Big] \epp
\end{align}
Finally it should be noticed that to recover the mirrored form factors one has take the complex conjugate and the reflection $x \to - x$
\begin{equation}
\langle \rho , \{   \mu_i^- \to   \mu_i^+ \}_{i=1}^n | \Psi^+(x) \Psi(0) | \rho \rangle = \Big( \langle \rho| \Psi^+(x) \Psi(0) | \rho  , \{   \mu_i^- \to \mu_i^+ \}_{i=1}^n \rangle^* \Big)\Big|_{x \to - x}     \epp
\end{equation}

\section{Fredholm determinant and Fredholm Pfaffian}\label{Fredholm}
Given a kernel $K$, a function $\rho$ defined on $\mathbb{R}$ and a domain $X \in \mathbb{R} $ we define a trace-class operator $K\rho$ satisfying   
\begin{equation}
\text{Tr}[K\rho]= \int_X dz K(z,z) \rho(z)  <  \infty \epp
\end{equation}
Its Fredholm determinant is defined as the infinite convergent sum
\begin{equation}\label{eqn:FD1}
\text{Det}( 1+  P_X K \rho P_X) = \sum_{n=0}^\infty \frac{1}{n!} \left( \prod_{j=1}^n \int_X dz_j \rho(z_j)  \right) \det_{i,j=1}^n K(z_i,z_j) \epc
\end{equation}
where we introduced the operator $P_X$ as a projector on the domain $X$ and the identity operator $1$. Given a second trace-class kernel $F \phi $ we can define the Fredholm determinant of the product of the two kernels $\text{Det}( 1+ F\phi K\rho)$ where the product is defined as
\begin{equation}
[F\phi K\rho](x,y) = \int dz F(x,z) \phi(z) K(z,y) \rho(y) \epp
\end{equation}
The Sylvester determinant theorem applies also to the Fredholm determinant for trace-class operators 
\begin{equation}
\text{Det}( 1+ F\phi K \rho) = \text{Det}( 1+ K \rho F\phi ) \epp
\end{equation} 

Following \cite{2008_Bornemann} we can see the Fredholm determinant as a discretization of the domain $X$ in $m$ points $\{ x_j \}_{j=1}^m$ with $m$ associated weights $\{ w_j \}_{j=1}^m$ such that given any integrable function $f(x)$ on $X$ we have
\begin{equation}
\int_X dy K(x_i,y) f(y) = \lim_{m \to \infty} \sum_{j=1}^m K(x_i, x_j) f(x_j) w_j \epc
\end{equation}
then the Fredholm determinant \eqref{eqn:FD1} is the limit of the determinant of the $m\times m$ matrix $\delta_{ij}+ [K\rho](x_i,x_j) $
\begin{equation}
\text{Det}( 1+  P_X K \rho P_X)  = \lim_{m \to \infty} \det_{i,j=1}^m \left( \delta_{ij} +[K\rho](x_i,x_j)  \right) \epp
\end{equation}

A generalization of the Fredholm determinant is given by the Fredholm Pfaffian defined for antisymmetric kernels. Given a two-by-two matrix kernel $ K_{ij}(u,v) = - K_{ji}(v,u)$
\begin{equation}
[\boldsymbol{K}\rho] (u,v) =  \begin{pmatrix}
 K_{11}(u,v) \rho(v) &    K_{12}(u,v) \rho(v) \\   K_{21}(u,v) \rho(v)   & K_{22}(u,v) \rho(v)
\end{pmatrix}  \epc
\end{equation} 
then we define its Fredholm Pfaffian on the domain $X$ as 
\begin{align}\label{eqn:FP1}
\text{Pf}& ( \boldsymbol{J}+  P_X \boldsymbol{K} \rho P_X) \\&
= \text{Pf}\left( \begin{pmatrix}
0  &     1 \\ -  1   & 0
\end{pmatrix}  +  \begin{pmatrix}
 P_X K_{11}  \rho P_X &     P_X K_{12}  \rho P_X  \\   P_X K_{21}  \rho P_X   &  P_X K_{22}  \rho P_X 
\end{pmatrix}  \right) 
\\&= \sum_{n=0}^\infty \frac{1}{n!} \left( \prod_{j=1}^n \int_X dz_j \rho(z_j)  \right) \text{Pf}{}_{i,j=1}^n \begin{pmatrix}
 K_{11}(z_i,z_j)   &    K_{12}(z_i,z_j)  \\   K_{21}(z_i,z_j)     & K_{22}(z_i,z_j)  
\end{pmatrix} \epc
\end{align}
where we introduced the two-by-two kernel
\begin{equation}
\boldsymbol{J} =  \begin{pmatrix}
0  &     1 \\ -  1   & 0
\end{pmatrix}  \epp
\end{equation}
The connection between Fredholm Pfaffian and Fredholm determinant is given by
\begin{equation}
\text{Det}( \boldsymbol{1} -  \boldsymbol{J}   \boldsymbol{K} ) = \text{Pf}( \boldsymbol{J} + \boldsymbol{K} )^2 \epp
\end{equation}
For more details on the Fredholm Pfaffian see \cite{1742-5468-2012-06-P06001,2007_Borodin}.

\section{Finite size expression of the field-field form factor in the Tonks-Girardeau regime}\label{Finite_Size}
We here compute the form factor for the field-field operator $\Psi^+(x) \Psi(0)$ (with $x>0$) at generic finite size $N$ by integrating the normalized Bethe wave function $\psi(\boldsymbol{x}| \boldsymbol{\lambda})$ at $c=\infty$ following the approach used in \cite{zvonarev}. We denote with $P$ a generic permutation of the whole set $[1,2, \ldots, N]$ and with $(-1)^{[P]}$ the sign of the permutation. The form factor is then given by
\begin{align}
& \langle \{ \lambda_i\}_{i=1}^N | \Psi^+(x) \Psi(0) |  \{\mu_i \}_{i=1}^N \rangle = \frac{N}{L^N} \Big[ \prod_{j=2}^N \int_0^L dx_j  \Big] \psi(\boldsymbol{x}| \boldsymbol{\mu})|_{x_1=0}   \: \psi^*(\boldsymbol{x}| \boldsymbol{\lambda})|_{x_1=x}  \\&
 = \frac{1}{L^N (N-1)!} \sum_{P,P'} (-1)^{[P]  + [P']}e^{-i x \lambda_{P_1}  } \int dx_2 \ldots dx_N \prod_{j=2}^N \text{sgn}(x_j - x)   e^{-i x_j (\lambda_{P_j} - \mu_{P'_j})} \\&
 = \frac{1}{L (N-1)!}  \sum_{P,P'} (-1)^{[P]  + [P']}e^{-i x \lambda_{P_1} } \prod_{j=2}^N  
q(\lambda_{P_j} - \mu_{P'_{j}}) \epc
\end{align}
where we defined 
\begin{equation}
q(\lambda) =- \frac{e^{-i L \lambda} + 1 - 2 e^{-i x \lambda} }{i L\lambda} = \frac{1}{L}\int_0^L  dz\: \text{sgn}(z - x) e^{-i z  \lambda} \epp
\end{equation}
We can now change permutation variable by defining a new permutation $P''$ such that $P'=P''P$ which allow then to sum over one set of permutations leading to
\begin{align}
& \langle \{ \lambda_i\}_{i=1}^N | \Psi^+(x) \Psi(0) |  \{\mu_i \}_{i=1}^N \rangle \\&
 = \frac{1}{L (N-1)!}  \sum_{P,P''} (-1)^{ [P'']} e^{-i x \lambda_{P_1} } \prod_{j=2}^N  
q(\lambda_{P_j} - \mu_{[PP'']_{j}}) \\
&= \frac{1}{L  }  \sum_{P''} (-1)^{ [P'']} \prod_{j=1}^N  
q(\lambda_j - \mu_{P''_{j}})  \sum_{k=1}^N \frac{e^{-i x \lambda_k } }{q(\lambda_k - \mu_{P''_k})}  \epc
\end{align}
which correspond to the expansion of the difference of two determinants
\begin{equation}
\langle \{ \lambda_i\}_{i=1}^N | \Psi^+(x) \Psi(0) |  \{\mu_i \}_{i=1}^N \rangle = \det_N \left( q(\lambda_a - \mu_b) + e^{-i \lambda_a x} \right)- \det_N  \left( q(\lambda_a - \mu_b) \right) \epc
\end{equation}
since the matrix $e^{-i \lambda_a x} $ is a rank one matrix. 

\newpage
\section*{References}

\bibliographystyle{iopart-num}
\bibliography{c_infty_particle_time_evolution}

\end{document}